\documentclass[11pt]{spieman}
\usepackage[a4paper,margin=1in]{geometry}
\usepackage{setspace}
\usepackage{tocloft}
\usepackage{lineno}
\usepackage{graphicx}
\usepackage{booktabs}
\usepackage{array}
\usepackage{amsmath,amssymb}
\usepackage{cite}
\usepackage{hyperref}
\usepackage[nameinlink,capitalise,noabbrev]{cleveref}
\usepackage{enumitem}
\usepackage{xcolor}

\newcommand{\Cn}{C_n^2}

\newcommand{\caseid}{\texttt{case\_id} {}}

\title{Open dataset for benchmarking scaling laws of high-energy laser atmospheric propagation}
\author[a,*]{Xusheng Xia}
\author[b,c]{Zhilin Xia} 
\affil[a]{School of Physics and Mechanics, Wuhan University of Technology, Wuhan, 430070, China} 
\affil[b]{State Key Laboratory of Advanced Glass Materials, Wuhan, 430070, China} 
\affil[c]{School of Material Science and Engineering, Wuhan University of Technology, 
Wuhan 430070, China} 

\date{}

\cftpagenumbersoff{figure}
\cftpagenumbersoff{table}

\begin{document}
\maketitle

\begin{abstract}
Scaling laws are increasingly used as fast surrogate models for high energy laser atmospheric propagation, yet their calibration and comparison still depend on large collections of high-fidelity wave-optics simulations. Existing studies usually rely on privately organized simulation outputs, which makes it difficult to reproduce published fits or evaluate new surrogate formulations on a shared benchmark. We present a public simulation dataset for high energy laser atmospheric propagation with coupled turbulence and thermal blooming. The release contains 226{,}500 cases spanning target speed, emission geometry, aperture diameter, visibility, aerosol model, beam quality, turbulence strength, and laser power. Data are organized as a case-level main table linked to indexed long-exposure irradiance arrays and centralized metadata, which supports statistical analysis without hiding the underlying field outputs. The simulation pipeline is based on split-step wave-optics propagation with turbulence, attenuation, and thermal-blooming models that have been validated against established propagation references. The dataset is intended for scaling-law calibration, benchmark comparison, surrogate-model training, sensitivity analysis, and inverse studies.


\end{abstract}


{\noindent \footnotesize\textbf{*}Xusheng Xia,  \linkable{xsxia@whut.edu.cn} }

\section{Introduction}

High-energy laser atmospheric propagation is governed by coupled diffraction, turbulence, attenuation, and thermal-blooming effects over long paths \cite{Gebhardt_Smith_1971,Gebhardt_1976,Smith_1977}. Wave-optics simulation remains one of the most reliable ways to resolve these interactions because it retains field-level structure rather than collapsing the problem into a few reduced metrics \cite{fleck1976timedependent,schmidt2010numerical,strohbehn1978laser}. This fidelity matters when the goal is not only to estimate a beam radius, but also to preserve long-exposure irradiance maps that can later support calibration, sensitivity analysis, and surrogate-model development.

Recent work shows renewed interest in fast predictive models for this problem, especially scaling laws that approximate wave-optics behavior over broad operating regimes \cite{Chen_2024,Hyde_Kalensky_Spencer_2024,Qiao_2010,
Van_2013,Xia_2026,Ye_2026,Zhang_2026}. These models are attractive because they replace repeated split-step propagation with compact formulas or lightweight regressors. Their usefulness, however, depends directly on the availability of trustworthy reference data. If the underlying simulation corpus is narrow, inconsistently organized, or inaccessible outside a single group, it becomes difficult to compare scaling-law proposals fairly or to diagnose where a surrogate fails.

The field still lacks a standardized public wave-optics benchmark for coupled turbulence and thermal blooming. Published studies often report fitted models or selected simulation results, but the full supporting datasets are usually private or bundled in formats that are difficult to reuse. As a result, researchers who want to test a new scaling law, benchmark a reduced-order predictor, or compare interpolation against extrapolation performance often have to reconstruct a large simulation corpus before they can start the actual modeling task.

To address this gap, we constructed and organized a public simulation dataset for high-energy laser atmospheric propagation under coupled turbulence and thermal blooming. The dataset comprises 226,500 cases and covers target speed, emission geometry, aperture diameter, visibility, aerosol model, beam quality, turbulence strength, and laser power across a parameter space designed for scaling-law development and comparison. The released data are organized as a case-indexed main table together with externally stored long-exposure irradiance arrays and centralized metadata, so that users can efficiently query operating conditions while retaining direct access to high-dimensional field outputs. Rather than serving only as a repository of simulation results, this dataset is intended as a common validation basis for scaling-law fitting, surrogate-model training, sensitivity analysis, and comparative benchmarking of fast predictive methods for high-energy laser propagation.

\section{Methods}

\begin{figure}[ht]
\centering
\includegraphics[width=0.92\textwidth]{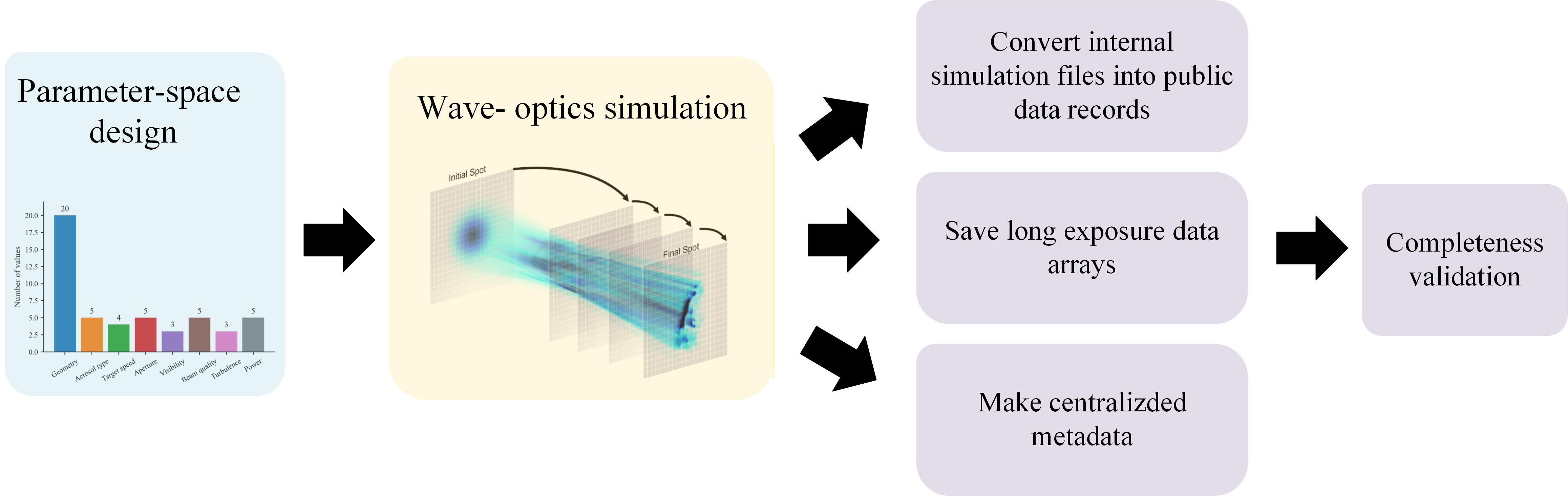}
\caption{Workflow of dataset generation, public-data conversion, and release validation.}
\label{fig:dataset_workflow}
\end{figure}

This section describes the end-to-end workflow used to generate, organize, and release the dataset. The workflow illustration is shown in \cref{fig:dataset_workflow}. We first introduce the wave-optics simulation framework used to produce the raw propagation outputs, and then detail the parameter-space design that yields 226,500 cases, with particular attention to conditional combinations arising from physical and application-oriented constraints rather than a naive Cartesian sweep. Raw propagation outputs are then generated using wave-optics simulations with turbulence and thermal-blooming modeling. These internal results are subsequently converted into a standardized public dataset structure consisting of case-level tables, external long-exposure irradiance arrays, and centralized metadata files. Finally, dataset integrity is verified through checks on case indexing, parameter coverage, file linkage, and completeness of the released records. 

\subsection{Simulation framework}

The released cases were generated with a split-step wave-optics model in which phase perturbations and free-space propagation are applied sequentially along the path \cite{frehlich2000simulation, charnotskii2013sparse}. The complex optical field $U(x,y,z)$ follows the paraxial propagation equation
\begin{equation}
\frac{\partial U(x,y,z)}{\partial z} = \frac{\textrm{i}}{2k}\nabla_{\perp}^{2} U(x,y,z),
\end{equation}
where $k$ is the optical wavenumber and $\nabla_{\perp}^{2}$ denotes the transverse Laplacian. In the split-step formulation, the field is sampled at discrete propagation planes. At each plane, turbulence and thermal-blooming phase terms are applied together with amplitude transmission, and the field is propagated over the next step by an angular-spectrum operator\cite{schmidt2010numerical}
\begin{equation}
U_i = Q_1(x,y)\mathcal{F}^{-1}\!\{Q_2(\kappa_x,\kappa_y) \, \mathcal{F}\!\left[Q_3(x,y)T_i U_{i-1} \exp(\textrm{i}\phi_T + \textrm{i}\phi_B)\right]\}.
\end{equation}
Here the subscript italic $i$ denotes the index of current phase screen; $\phi_T$ and $\phi_B$ denote the turbulence and thermal-blooming phase screens; $T_i$ is the amplitude transmission between adjacent planes; $Q_1$, $Q_2$ and $Q_3$ are the frequency-domain propagation phase factors.

Atmospheric turbulence is represented with a modified von K\'arm\'an spectrum. The high-frequency phase screen is generated by spectral synthesis,
\begin{equation}
\phi_H(x,y)=\mathcal{F}^{-1}\!\left[\Phi_{\phi}(\kappa)^{1/2} g(\kappa)\right],
\end{equation}
where $g(\kappa)$ is a complex Gaussian random field and $\kappa=\sqrt{\kappa_x^2+\kappa_y^2}$. The spectrum is written in modified von K\'arm\'an form as\cite{lu2024simulation}
\begin{equation}
\Phi_{\phi}(\kappa)=0.033\, C_n^2 \left(\kappa^2+\kappa_0^2\right)^{-11/6}
\exp\!\left(-\frac{\kappa^2}{\kappa_m^2}\right).
\end{equation}
Low-frequency components are then restored through a Zernike-based compensation term\cite{jianzhu2012methods},
\begin{equation}
\phi_L(x,y)=\sum_{i,k} R_{ik} g_i Z_k(x,y),
\end{equation}
and the full turbulence phase screen is $\phi_T=\phi_H+\phi_L$. This treatment keeps the released data aligned with accepted long-exposure propagation statistics \cite{andrews2005random,charnotskii2013sparse}. It is also compatible with established adaptive-optics simulation practice, although the present released dataset does not contain adaptive-optics control \cite{townson2019aotools}.

Thermal blooming is treated through a Fourier-domain solution of the heating-induced refractive-index perturbation, following the standard coupling between absorbed laser intensity, advection, and diffusion \cite{fleck1976timedependent,strohbehn1978laser,schleijpen2024thermal}. In the spectral domain, the heating-driven refractive-index field evolves as
\begin{equation}
\hat n_b(\kappa,z,t+\Delta t)=e^{(-\textrm{i}\kappa\cdot v-\eta \kappa^2)\Delta t}\hat n_b(\kappa,z,t)
+ \Gamma \alpha_{\mathrm{abs}}\!\int_{t}^{t+\Delta t}
e^{(\textrm{i}\kappa\cdot v+\eta \kappa^2)(t'-t-\Delta t)} \hat I(\kappa,z,t')\,\mathrm{d}t',
\label{bloom}
\end{equation}
where $v$ is the transverse wind vector, $\eta$ is the thermal diffusivity, $\Gamma$ is the thermo-optic coefficient of air, $\alpha_{\mathrm{abs}}$ is the absorption coefficient, and $\hat I$ is the Fourier transform of the laser intensity. The corresponding thermal-blooming phase screen is
\begin{equation}
\phi_B(x,y)=2\pi k \int \mathcal{F}^{-1}\!\left[\hat n_b(\kappa,z)\right]\mathrm{d}z.
\end{equation}
The resulting thermal phase is accumulated along the path and combined with the turbulence phase at each split-step update. Atmospheric attenuation enters the simulation through amplitude transmission terms, so the released dataset preserves both distortive and absorptive effects rather than only a phase-only approximation.

Target motion is included at the scenario-definition level through the target-speed parameter and will be considered as pseudo-wind-speed in thermal blooming \cref{bloom}. For dataset generation, the default numerical configuration uses phase screens sampled on a ($128\times128$) grid, with 50 phase screens distributed along the propagation path, and long-exposure irradiance obtained by averaging 100 transient propagation realizations. This baseline discretization is adopted for most cases in the dataset. For scenarios with stronger thermal blooming or stronger turbulence, however, the phase-screen sampling resolution is increased to ensure that the phase difference between adjacent pixels does not exceed $\pi$, thereby maintaining adequate numerical representation of the phase distortion. Each released array therefore corresponds to a long-exposure irradiance map generated under a numerically controlled wave-optics setting, while the public dataset records the case-level scenario parameters together with the linkage required to retrieve the associated long-exposure array.

\subsection{Parameter-space design}

\begin{figure}[htb]
\centering
\includegraphics[width=\textwidth]{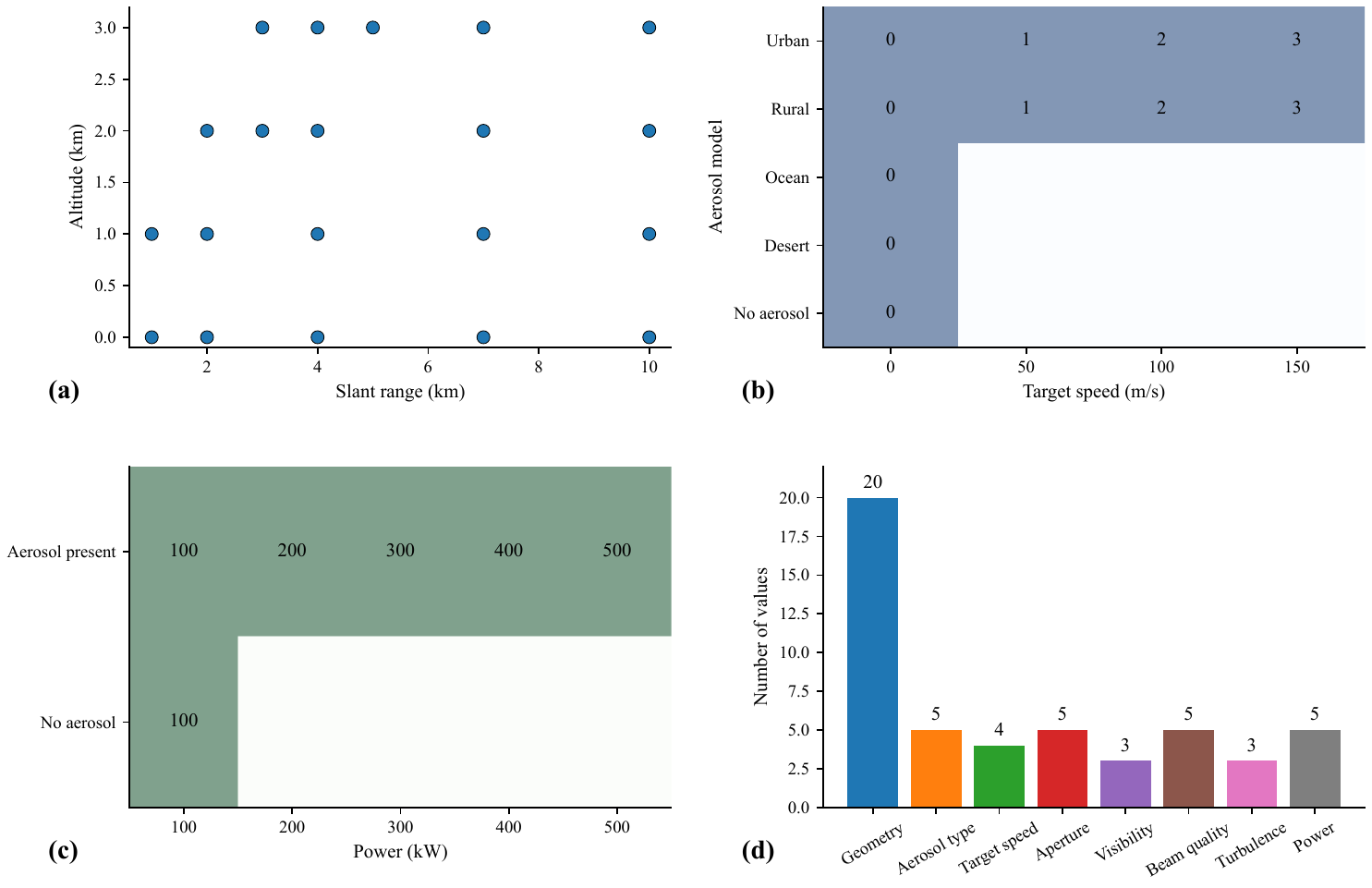}
\caption{Summary of the parameter-space design. The geometry points in (a) show the span of the 20 altitude--slant-range combinations, while the eligibility matrices in (b) and (c) highlight the conditional combinations that prevent the release from becoming a naive full Cartesian product. (d) provides a brief summary of the parameter-space cardinalities.}
\label{fig:parameter_space}
\end{figure}

The parameter space was designed to support surrogate-model calibration and comparison rather than to enumerate every mathematically possible combination. The release spans four target speeds (0, 50, 100, and 150 m/s), 20 altitude--slant-range geometries, five aperture diameters (100--500 mm), three visibility levels (5, 15, and 23 km), five beam-quality factors ($\beta=3, 4, 6, 8, 10$), three turbulence strengths ($\Cn = 5\times10^{-15}, 1\times10^{-15}, 5\times10^{-16}$), and up to five power settings (100--500 kW). Other parameters were designed as fixed. The design logic is summarized in \cref{tab:parameter_space,fig:parameter_space}.

\begin{table}[htb]
\centering
\caption{Parameter-space design and case-count logic used for the released dataset.}
\footnotesize
\label{tab:parameter_space}
\begin{tabular}{p{4.5cm}p{4.1cm}p{5.5cm}}
\toprule
Parameter & Values & Conditional rule \\
\midrule
Target speed (m/s) & 0, 50, 100, 150 & Ocean, desert, and no-aerosol conditions are defined only for the static target subset (0 m/s). \\
Emission geometry & 20 combinations & Slant range spans 1--10 km and altitude spans 0--3 km in scenario-specific pairings. \\
Aperture diameter (mm) & 100, 200, 300, 400, 500 & Applied across all eligible scenarios. \\
Visibility (km) & 5, 15, 23 & Visibility is not distinguished for no-aerosol scenes. \\
Aerosol model & Urban, Rural, Ocean, Desert, No-aerosol & Urban and Rural are broadly combined with dynamic targets; Ocean, Desert, and No-aerosol are static-target only. \\
Beam quality ($\beta$) & 3, 4, 6, 8, 10 & Applied across all eligible scenarios. \\

Laser power (kW) & 100, 200, 300, 400, 500 & For no-aerosol scenes only 100 kW is used because thermal blooming is inactive. \\
Laser wavelength (nm) & 1064 \\
Laser spot radius & 1/3$\times$ aperture diameter \\
Atmosphere turbulence profile & HV modified model \\
$A$ in HV model (m$^{-2/3}$) & $5\times10^{-15}$, $1\times10^{-15}$, $5\times10^{-16}$  & Applied across all eligible scenarios. \\
Wind speed profile & Bufton model\\

\midrule
Case counts  & Total: 226{,}500 cases. \\
\bottomrule
\end{tabular}
\end{table}

The aerosol absorption and scattering coefficients are calculated from the listed visibilities and models using LOWTRAN \cite{kneizys1988users}. The wind speed model is Bufton wind profile:
\begin{equation}
v_W\mathrm{\mathrm{=5\ +}\ 30}\times \exp{\left[-\left(\frac{h-9400}{4800}\right)^2\right]},
\end{equation}
where the result   is wind speed in m/s,   is altitude (in meters). The turbulence profile is Hufnagel-Valley model \cite{andrews2009near}
\begin{equation}
C_n^2\left(h\right)=0.00594\left(\frac{v_W}{27}\right)^2\left(10^{-5}h\right)^{10}\textrm{e}^{-h/1000}+2.7\times10^{-16}\textrm{e}^{-h/1500}+A\textrm{e}^{-h/100},
\label{eq:hv}
\end{equation}
where the result is in m$^{-2/3}$,  $h$ is altitude (in meters), and  $v_W$ is wind speed at high altitude (in m/s, here using Bufton profile),  $A$ is a parameter to control near surface turbulence strength. The values of  $A$ are listed in the table.

To avoid generating large amounts of redundant or weakly informative data, the parameter space was not constructed as a naive full Cartesian product. Instead, several combinations were selectively excluded when they were not expected to provide additional physical regimes or benchmark value. This design was intended to keep the dataset large and representative while reducing unnecessary storage and duplication of cases whose propagation behavior would be effectively equivalent.

Two conditional rules are particularly important in this context. First, the aerosol categories Ocean, Desert, and No-aerosol were included only for the static-target subset, whereas the dynamic-target subset was restricted to the Urban and Rural scenarios. The main motivation is that the study of moving targets in this dataset is closely tied to thermal-blooming-related effects. In oceanic and desert aerosol conditions, however, aerosol absorption is comparatively weak, so thermal blooming is less pronounced and the added value of introducing target motion is limited. By contrast, Urban and Rural atmospheres provide stronger and more practically relevant thermal-blooming conditions, making them more suitable for dynamic-target cases and giving broader benchmark coverage for motion-related propagation studies. Second, laser power was not swept under the no-aerosol condition. In the absence of thermal blooming, the remaining propagation effects in this dataset—primarily turbulence and extinction—are linear with respect to optical power. Under such conditions, changing the transmitted power does not alter the normalized propagation outcome, and repeated power settings would therefore increase dataset size without introducing a distinct physical regime. For this reason, power scanning was reserved for conditions in which thermal blooming is active and power becomes a physically meaningful control parameter.

The resulting counts are 91{,}500 cases for the static-target subset and 135{,}000 cases for the dynamic-target subset, for a total of 226{,}500 cases. This coverage is broad enough to support interpolation studies within common operating regions and extrapolation studies across geometry, turbulence, and blooming strength. At the same time, the design remains physically interpretable because each conditional omission has a modeling reason that can be stated explicitly.

\subsection{Case generation and indexing}

The public release was produced by converting private simulation outputs into a case-indexed dataset. Each simulation scene was assigned a persistent identifier (\caseid) and then mapped to the corresponding long-exposure irradiance array through explicit linkage fields. This design decouples high-dimensional field storage from scenario metadata: users can filter or stratify the case table first and load arrays only for the cases they need.

During conversion, the generation pipeline normalized parameter naming, attached release-level version information, and created indexing fields that make the mapping from a table row to an array file deterministic. A row in the case table therefore functions as a compact scene manifest: it stores the scenario variables, identifies the subset or release version, and provides the key needed to retrieve the long-exposure array. 

This indexing strategy was chosen to support several common workflows. Researchers fitting scaling laws may only need the tabular variables and selected derived beam metrics. Researchers training image-based surrogates can use the same table to define splits and then retrieve the linked arrays. Because the identifiers are stable, benchmark definitions can be shared without copying the full dataset or relying on private bookkeeping conventions.

\section{Data Records}

The public dataset is deposited in Zenodo under DOI \href{https://doi.org/10.5281/zenodo.19157773}{10.5281/zenodo.19157773}. The release contains a case-level main table, metadata files, a variable dictionary, and the associated array files for long-exposure irradiance distributions. The repository-level organization is summarized in \cref{tab:data_organization,tab:variable_summary}. The metadata folder explicitly records that the current public release is version \texttt{v1.0}, that it belongs to the \texttt{general} subset, that it contains 226{,}500 public cases, and that the packaged public dataset originates from 11{,}325 source files.

\begin{table}[htb]
\centering
\caption{Public data organization in the Zenodo release.}
\footnotesize
\label{tab:data_organization}
\begin{tabular}{p{2.8cm}p{4.6cm}p{6.8cm}}
\toprule
Component & Path / format & Description \\
\midrule
Dataset summary & \path|metadata/dataset_info.json| & Release-level descriptor listing version (\texttt{v1.0}), subset name (\texttt{general}), total case count (226{,}500), number of source files (11{,}325), the main table path, and the external-array mapping rule. \\
Variable dictionary & \path|metadata/variable_dictionary.csv|    & Field-level schema for all public variables, including group, data type, units, semantic role, text description, and recommended storage location. \\
Case-level main table & \path|tables/cases.csv| & Tabular record with one row per simulation case. Stores scalar inputs, selected derived outputs, and the linkage needed to recover externally stored long-exposure arrays. \\
Extinction table & \path|extinct/*.csv| & Scattering, extinction and absorption profile for different aerosol types and different visibilities. \\
Long-exposure arrays & \path|long_exposure_{case_id_range}.h5:/long_exposure/{case_id}| & HDF5 container of target-plane long-exposure irradiance matrices. Each dataset is keyed by \texttt{case\_id}, which avoids ambiguity when table rows are filtered or reordered. \\

\bottomrule
\end{tabular}
\end{table}


\begin{table}[t]
\centering
\caption{Representative variable groups and key fields for dataset reuse.}
\label{tab:variable_summary}
\footnotesize
\begin{tabular}{p{3.0cm}p{3.0cm}p{7.0cm}}
\toprule
Group & Representative fields & Role in reuse workflows \\
\midrule
Scene identity & \path|case_id|, \path|dataset_version|, \path|subset| & Unique indexing and split management for benchmark protocols and release tracking. \\
Geometry and motion & \path|target_distance_km|, \path|target_height_km|, \path|target_speed_mps| & Defines propagation path and dynamic regime for interpolation or extrapolation studies. \\
Beam and aperture & \path|aperture_diameter_mm|, \path|beam_quality_beta0|, \path|power_kw| & Governs source-side operating conditions for scaling-law fitting and sensitivity analysis. \\
Atmosphere and turbulence & \path|visibility_km|, \path|cn2_earth_reference|, \path|outer_scale_L0| & Encodes attenuation and turbulence regime information used by the simulator. \\
Array linkage & \path|case_id|, \path|target_irradiance_long_exposure|, \path|target_plane_extent_m| & Maps each table row to the external HDF5 irradiance array and its spatial extent. \\
Derived beam metrics & \path|spot_r0632_cm|, \path|spot_r084_cm|, \path|transmission| & Supports filtering, regression, and evaluation before loading the full irradiance matrix. \\
\bottomrule
\end{tabular}
\end{table}


At the repository level, the main table \path|tables/cases.csv| acts as the primary entry point. Each row in this table represents one simulated case and stores the scalar input parameters, selected derived quantities, and the linkage needed to recover externally stored arrays. Integers, float-pointing numbers and strings are directly recorded, while small matrices are recorded in JSON format.

Large field outputs are not embedded directly in the case table. Instead, the \caseid maps the long-exposure irradiance variable, \path|target_irradiance_long_exposure|, to the external HDF5 location \path|long_exposure_{case_id_range}.h5:/long_exposure/{case_id}|. In practice, this means that the case identifier is the key that binds the scene-level row in \path|cases.csv| to the corresponding target-plane irradiance matrix in the HDF5 container.

The variable dictionary is released in both CSV and JSON form. Each entry records the public variable name, target group, data type, units, semantic role, human-readable description, and recommended storage mode. The schema is organized into four practical groups. The \textit{inputs} group includes atmospheric and propagation settings such as \path|visibility_km|, \path|target_distance_km|, \path|target_height_km|, \path|target_speed_mps|, \path|power_kw|, and the phase-screen and time-step parameters. The \textit{outputs} group contains the long-exposure irradiance array. The \textit{grid} group records the target-plane extent used to interpret the external array. The \textit{derived} group provides reusable beam descriptors such as centroid coordinates, second-moment radii, encircled-energy radii, power-density measures, total received power, transmission, and coherence length.

For reuse, the most important fields are the stable case identifier, the geometry and motion variables, the beam and atmosphere descriptors, and the key used to locate the long-exposure array. These fields support both tabular workflows and array-based workflows. A user can, for example, identify all cases for a given turbulence regime in the main table, inspect tabulated derived metrics such as \path|spot_r0632_cm| and \path|transmission|, and then load only the linked irradiance arrays needed for surrogate-model training. Because the linkage is explicit rather than positional, the release remains interpretable even when files are moved, mirrored, or subsetted for a benchmark.

\section{Technical Validation}

Technical validation is performed at two levels: validation of the underlying propagation model and validation of the public data release. The model-level checks reuse established comparisons from the same simulation framework, while the release-level checks verify that the public package is complete, internally consistent, and suitable for downstream reuse.

\subsection{Model-level Validation}
\begin{figure}[htb]
\centering
\includegraphics[width=0.7\textwidth]{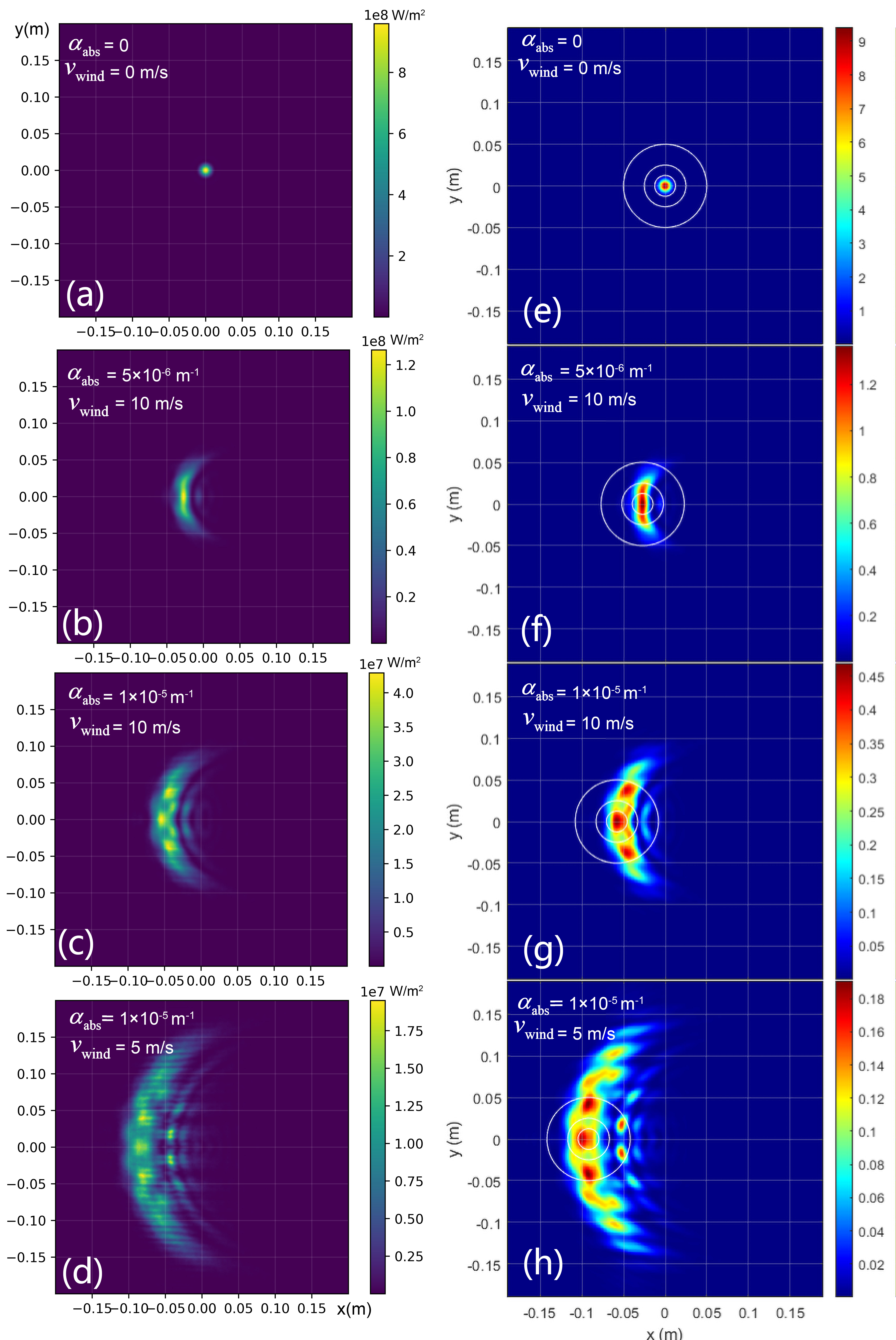}
\caption{Resultant beam intensity distribution in steady-state thermal
blooming simulation for a 200-kW laser beam with 500 mm aperture and 5
km focus distance. Left column: intensity spots computed by the proposed
method; right column: results reproduced from Schleijpen\cite{schleijpen2024thermal} for
comparison.
}
\label{fig:bloom_compare}
\end{figure}

At the model level, the wave-optics framework was checked against reference calculations for both thermal blooming and turbulence-sensitive long-exposure propagation.

For thermal blooming validation, we simulate a steady-state propagation
scenario with a transmitted power of 200 kW, an aperture diameter of 500
mm, and a focus distance of 5 km. The atmospheric absorption coefficient
${\alpha_{{\textrm{abs}}}}$ is set to $5 \times {10^{ - 6}}$ m$^{-1}$
and $1 \times {10^{ - 5}}$ m$^{-1}$, while ${v_{{\textrm{wind}}}}$ is
set to 5 m/s and 10 m/s. The resulting beam intensity profiles are shown
in \cref{fig:bloom_compare}. A comparison with reference \cite{schleijpen2024thermal}
simulations under
identical conditions shows good agreement in both the beam shape and
peak intensity. \cref{fig:bloom_compare} clearly illustrates the progressive distortion of
the beam due to thermal blooming effects. In the baseline case (\cref{fig:bloom_compare}a, e
), where both absorption and wind velocity are zero, the beam retains
its near-diffraction-limited focus. As thermal blooming
intensifies---driven by increased absorption coefficients and wind
velocities---the beam becomes increasingly asymmetric, stretched in the
wind direction, and exhibits reduced peak intensity. \cref{fig:bloom_compare}b, f show
that a moderate absorption with a wind velocity of 10 m/s leads to
lateral beam shifting and intensity spreading. This trend becomes more
pronounced in \cref{fig:bloom_compare}c, d and g, h, where the higher absorption and
smaller wind speed induces severe thermal lensing and results in
increasing intensity profiles. Overall, our results (left column) are in
good agreement with those of Schleijpen\cite{schleijpen2024thermal} (right column), both in
structure and intensity scale, confirming the validity of the proposed
simulation approach.

\begin{figure}[htb]
\centering
\includegraphics[width=0.7\textwidth]{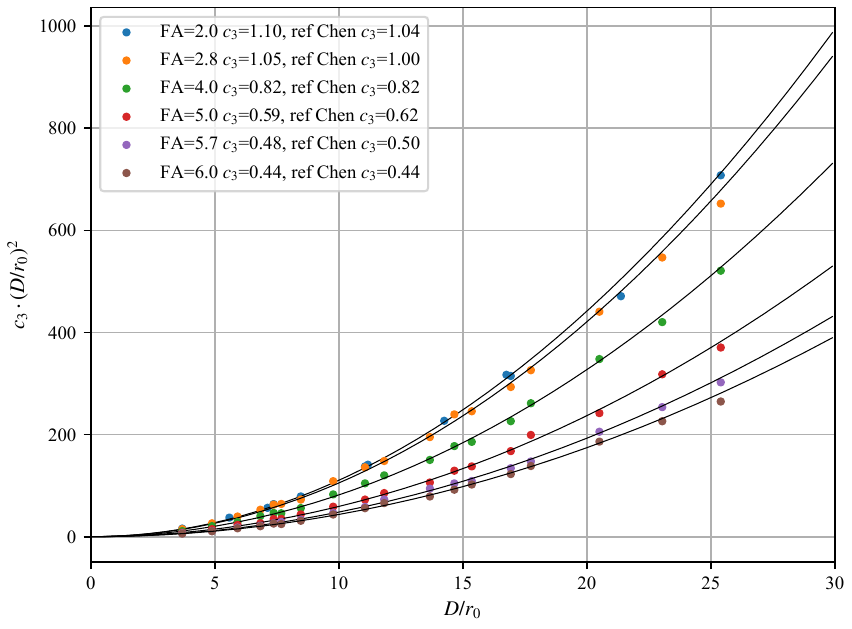}
\caption{Fitting results of long-exposure beam propagation using the Yura’s scaling law for truncated Gaussian beams. Data corresponds to simulations with apertures of $D =$ 0.5, 1.0, and 1.5 m, transmission distances $L$ between 5 and 20 km, and propagation altitude of 5 km under HV57 turbulence conditions.
}
\label{fig:turb}
\end{figure}

For turbulence validation, the widely used Hufnagel-Valley 5/7 (``HV5/7")
turbulence is applied in simulation. The general HV model is expressed
as \cref{eq:hv}. The shorthand ``HV 5/7"
refers to a commonly used configuration of this model where $v =
21$ m/s, $A = 1.7 \times {10^{ - 14}}{{\textrm{ m}}^{ - 2/3}}$.
Long-exposure simulations are performed under HV5/7 conditions for
transmitter apertures $D = 0.5$, 1.0, and 1.5 m, propagation distances
$5 \le L \le 20$ km, and a propagation height of 5 km. The far-field
beam widths are fitted using the Yura's scaling law for truncated
Gaussian beams \cite{Chen_2024}, given by:
\begin{equation}
a_{Y}^{2}={{\left( {{c}_{\text{a}}}{{F}_{\text{a}}}\frac{\lambda L}{D} \right)}^{2}}\left[ 1+{{c}_{3}}{{\left( \frac{D}{{{r}_{0}}} \right)}^{2}} \right],
\end{equation}
\begin{equation}
{{c}_{\text{a}}}=1.26\exp (-1.04{{F}_{\text{a}}})+0.16.
\end{equation}
Here, ${a_Y}$ is the 63.2\% encircled energy radius at the target
plane, and $F_{\text{a}}$ is the truncation ratio, defined as the ratio of
the aperture diameter to the beam waist radius, $D/{a_0}$. 
For each aperture diameter $D$, we selected corresponding ${a_0}$ values to yield ${F_a} =
D/{a_0} =$ 2.0, 2.8, 4.0, 5.0, 5.7 and
6.0, respectively. These $F_{\text{a}}$ values of were deliberately selected
to match those used in Chen's work \cite{Chen_2024}, in order to facilitate direct
comparison of accuracy. As shown in \cref{fig:turb}, the fitted values of the
scaling parameter ${c_3}$ under different truncation conditions are in
close agreement with the reference results reported in Chen's work\cite{Chen_2024}.

\subsection{Dataset-level Validation}
At the dataset level, the release was checked against the designed parameter-space logic before packaging. The checks include verification of the total case count and the subset counts, uniqueness of stable identifiers, consistency between table-level linkage fields and array-level storage, and scans for malformed or missing public-format records. Shape, type, and unit conventions were also harmonized so that the same field has the same interpretation across the release.

\section{Usage Notes}

The dataset is intended primarily for scaling-law calibration, validation, and comparison. A typical workflow starts from the case-level table: users filter the cases of interest by geometry, turbulence strength, visibility, or aerosol model, define a split, and then load the linked long-exposure arrays for the selected cases. This separation makes it straightforward to benchmark tabular surrogates, image-based surrogates, or hybrid models that combine both views of the same case.

A minimal usage pattern is to read the main table, select a row by \caseid, and then use the linkage fields in that row to open the corresponding long-exposure array file. The table provides the scenario context, while the array stores the full irradiance distribution. This pattern also supports benchmark reproducibility: once a split is defined as a list of case identifiers, other researchers can reconstruct the exact benchmark without relying on local file ordering.

The release is especially suitable for benchmark settings that distinguish interpolation from extrapolation. For example, a user may train on a subset of target speeds or turbulence strengths and then evaluate on held-out regimes. Another option is to reserve an entire geometry band or aerosol category for testing. Because the parameter space is broad but structured, such splits can probe generalization without obscuring the physical interpretation of the failure modes.

Two benchmark protocols are likely to be useful from the start. An interpolation protocol can sample all parameter dimensions but hold out a random subset of \caseid values within each regime, which tests how well a surrogate fits the published domain. An extrapolation protocol can hold out one regime family entirely, such as the strongest turbulence level or the longest geometry band, which tests whether the surrogate preserves physically meaningful trends outside the calibration subset. In both cases, publishing the list of case identifiers is enough to make the benchmark reproducible.

Users should keep three caveats in mind. First, the dataset is simulation-based and therefore reflects the assumptions of the underlying wave-optics model. Second, the coverage is finite: although broad, it does not represent every atmospheric or platform condition of practical interest. Third, some dimensions are conditionally defined rather than uniformly crossed, so benchmark protocols should respect the design logic summarized in \cref{tab:parameter_space,fig:parameter_space} rather than treating absent combinations as missing data errors.

\section*{Data and code availability}

All data generated and described in this study are publicly available in Zenodo under DOI \href{https://doi.org/10.5281/zenodo.19157773}{10.5281/zenodo.19157773}. The repository contains the complete simulation dataset, including the case-level tabular records, metadata files, variable dictionary, and the associated array files for long-exposure irradiance distributions. The dataset is distributed under the Creative Commons Attribution 4.0 International license. The simulation and data-conversion codes used in this study cannot be shared at this time because the code also forms part of an ongoing paper.


\section*{Funding}
The research was funded by the National Natural Science Foundation of China (No. 62505236) and the Fundamental Research Funds for the Central Universities (No. 104972026RSCbs0126).

\section*{Author Contributions}
X.X. conceived the research, developed the methods, generated the dataset and wrote the manuscript. Z.X. supervised the research and contributed to the manuscript revision.

\section*{Disclosures}
The authors declare no competing interests.

\noindent AI tool Chat-GPT 5.4 was used to improve language and grammar.

\bibliographystyle{spiejour}
\bibliography{references.bib}


\vspace{2ex}\noindent\textbf{Xusheng Xia} is an assistant researcher at Wuhan Univeristy of Technology. He received his BS degree in chemistry from Nanjing Univeristy in 2014, and his PhD degree in chemistry from Dalian Institute of Chemical Physics in 2020. His current research interests include laser atmospheric propagation, adapted optics and optical films.

\vspace{1ex}
\noindent Biographies of the other authors are not available.

\listoffigures
\listoftables

\end{document}